\documentclass[aps,prl,reprint,twocolumn,superscriptaddress,amsmath,amssymb]{revtex4-2}
\usepackage{graphicx,bm,mathptmx,physics,color,anyfontsize}
\usepackage[colorlinks=true,allcolors=blue]{hyperref}
\usepackage[sort&compress]{natbib}
\def\red#1{{\color{black}#1}}

\begin{document}
\title{Over-Barrier Ionization Dynamics Studied by Backpropagation}

\author{Yongzhe~Ma}
\affiliation{State Key Laboratory of Precision Spectroscopy, East China Normal University, Shanghai 200241, China}
\affiliation{College of Science, Harbin Institute of Technology, Weihai, Shandong 264209, China}
\author{Qingcao~Liu}
\affiliation{College of Science, Harbin Institute of Technology, Weihai, Shandong 264209, China}
\author{Hongcheng~Ni}
\email{hcni@lps.ecnu.edu.cn}
\affiliation{State Key Laboratory of Precision Spectroscopy, East China Normal University, Shanghai 200241, China}
\affiliation{Collaborative Innovation Center of Extreme Optics, Shanxi University, Taiyuan, Shanxi 030006, China}
\author{Jian~Wu}
\email{jwu@phy.ecnu.edu.cn}
\affiliation{State Key Laboratory of Precision Spectroscopy, East China Normal University, Shanghai 200241, China}
\affiliation{Collaborative Innovation Center of Extreme Optics, Shanxi University, Taiyuan, Shanxi 030006, China}
\affiliation{Chongqing Key Laboratory of Precision Optics, Chongqing Institute of East China Normal University, Chongqing 401121, China}

\begin{abstract}
Tunneling and over-barrier ionization are the primary processes of strong-field ionization of atoms and molecules. While the dynamics of tunneling ionization have been extensively studied, exploration of over-barrier ionization dynamics has remained a significant challenge. In this study, we investigate the dynamics of over-barrier ionization using the backpropagation method specifically adapted for this context. By analyzing the topology of the backpropagating trajectories, we differentiate the contributions of tunneling and over-barrier ionizations to the distributions of photoelectron momentum and ionization time. While the transition from tunneling to over-barrier ionization is known to depend on the field strength, our results reveal that it is also influenced by the initial transverse momentum of the outgoing electron. We clarify how ionization probabilities vary with intensity for each mechanism, highlighting a competitive relationship between them. We further find that accounting for the Stark shift is essential for accurately determining the threshold field strength for over-barrier ionization. Our work provides a detailed understanding of the dynamics of over-barrier ionization and lays the groundwork for exploring new mechanisms in intense laser-matter interactions.
\end{abstract}

\maketitle

\begin{figure}[b]
	\centering
	\includegraphics[width=\columnwidth]{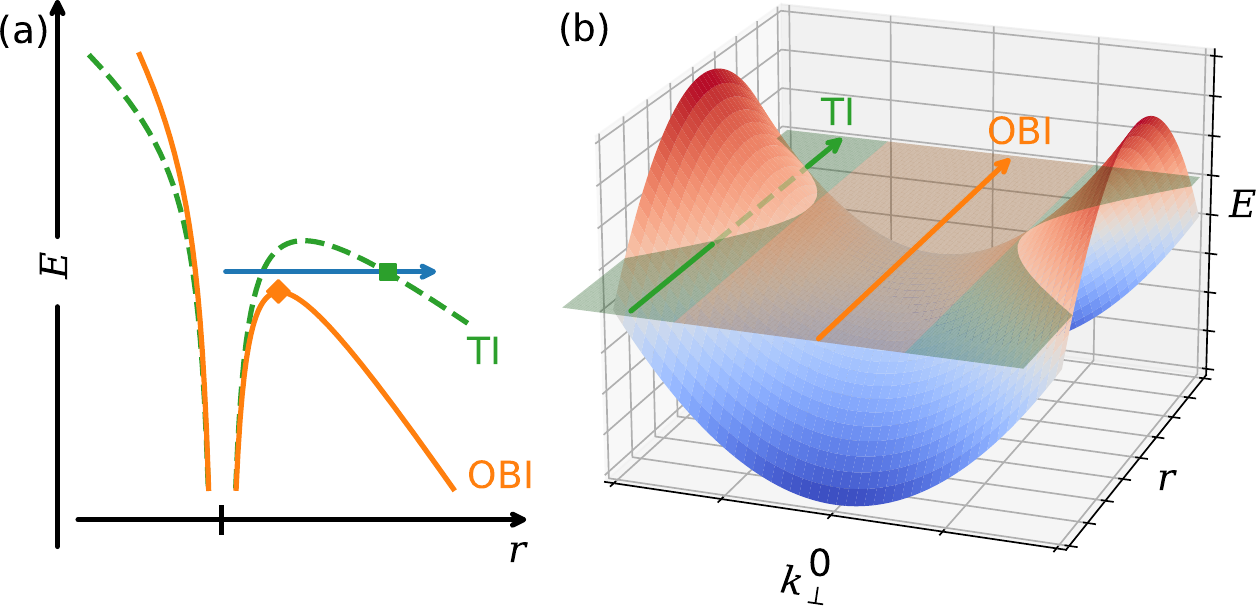}
	\caption{(a) Sketch of tunneling ionization (TI) and over-barrier ionization (OBI). (b) Effective potential barrier as a function of radial position $r$ and initial transverse momentum $k_\perp$. Green and orange arrows represent TI and OBI, respectively.}
	\label{fig:potential}
\end{figure}

The interaction of intense laser fields with matter produces a variety of intriguing physical phenomena, such as above-threshold ionization \cite{Agostini1979,Eberly1991,Becker2002}, nonsequential double ionization \cite{Fittinghoff1992,Walker1994}, and high-order harmonic generation \cite{Ferray1988,Krause1992,Macklin1993,Corkum1993,Lewenstein1994,Popmintchev2010}. These phenomena can be explained by different ionization mechanisms that are characterized by the Keldysh parameter $\gamma = \frac{\omega}{F}\sqrt{2I_p}$, where $I_p$ represents the ionization potential, $\omega$ is the angular frequency, and $F$ is the field strength of the laser pulse. When $\gamma \gg 1$, the laser pulses have high frequencies and relatively low field strengths, allowing atoms and molecules to absorb multiple photons, leading to ionization through the multiphoton ionization (MPI) mechanism. Conversely, when $\gamma \ll 1$, the laser pulses have low frequencies and high field strengths, which enables bound electrons to tunnel through the potential barrier or go over it to be ionized, known respectively as tunneling ionization (TI) and over-barrier ionization (OBI), as shown in Fig.~\ref{fig:potential}(a). Both MPI and TI have been extensively studied over the past few decades \cite{Keldysh1965,Ammosov1986,Ivanov2005,Popov2004,Popruzhenko2014,Keldysh2017,Ma2024}.

For OBI, a simple model \cite{Augst1989} establishes the threshold field strength $F_{\rm th} = {I_p^2}/{(4Z)}$, where $Z$ is the charge of the residual ion, and has effectively explained ion production rates observed in experiments \cite{Augst1991,Auguste1992}. To improve the accuracy of this threshold condition, it is often necessary to account for the Stark shift in the binding energy \cite{Bauer1997,Gorlinger2000}. It is important to note that the OBI mechanism has distinct characteristics compared to TI. For instance, the ionization rate formula effective in the TI regime cannot be directly applied to the OBI regime due to its tendency to significantly overestimate the observed ionization rates. To address this, methods based on strong-field approximation (SFA) \cite{Krainov1995,Krainov1997,Delone1998,Klaiber2024} and empirical formulas \red{\cite{Bauer1999,Scrinzi1999,Tong2005,Zhang2014,Bauer2017,Remme2025}} are employed to assess the ionization rate in the OBI regime. However, these methods face a significant limitation in their inability to provide detailed descriptions of electron dynamics within the OBI regime.

Initiating and manipulating electron motion in atoms and molecules is a central goal in ultrafast physics. Current research on OBI primarily focuses on ionization probability, excitation probability and photoelectron momentum distribution (PMD) \cite{Yang2012,Bauer2014,Song2014,Jheng2018}. Furthermore, alkali-metal atoms and large polyatomic molecules, owing to their lower ionization potentials, serve as model systems for studying laser-induced electron diffraction \cite{Belsa2022} and single-photon ionization \cite{Ma2023} in the OBI regime. However, due to the Gaussian-shaped temporal intensity profile of the laser pulses, there is a gradual transition in the ionization mechanism from TI to OBI. Precisely identifying and differentiating the individual contributions of these two mechanisms is a significant challenge that requires thorough exploration. Up to now, the dynamic process of OBI has remained unclear. This is largely because the prevailing theories derived from SFA assumes a zero-range potential. In this scenario, a triangular potential barrier is always present, regardless of the intensity of the laser field, thus limiting the study to TI alone.

In this Letter, we pioneer an investigation into the electron dynamics within the OBI regime, employing the backpropagation method specifically tailored for this regime. Our findings elucidate the dynamic differences between electrons experiencing OBI and those in the TI regime, and we clearly distinguish the respective contributions of these two ionization mechanisms to the PMD and the distribution of ionization time. Our study reveals that the occurrence of OBI depends not only on the field strength but also on the initial transverse momentum of the electron. We delve into the intensity-dependent ionization probabilities for both TI and OBI, revealing a competitive interplay between these mechanisms. We determine that the influence of Stark shift on the initial-state energy is crucial for accurately assessing the threshold field strength.

\begin{figure}[b]
	\centering
	\includegraphics[width=\columnwidth]{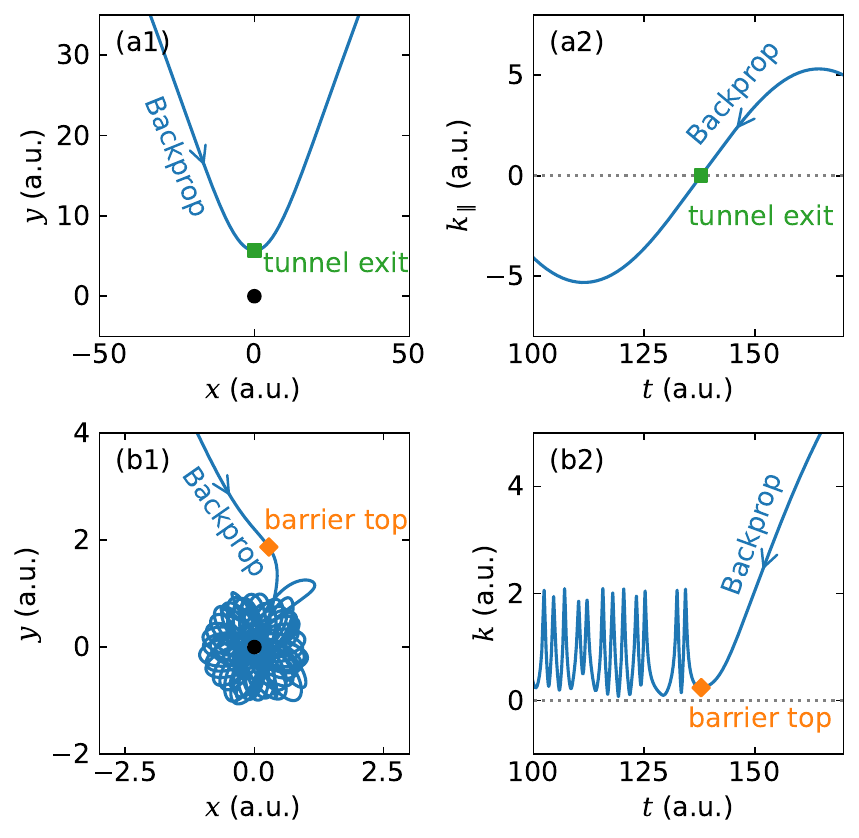}
	\caption{(a1) Typical TI trajectory and (a2) corresponding time dependence of the electron velocity $k_\parallel$ in the instantaneous field direction. (b1) Typical OBI trajectory and (b2) corresponding time dependence of the electron speed $k$. The arrow indicates the direction of the backpropagating trajectory. The green square and orange diamond represent the tunnel exit and the barrier top, respectively.}
	\label{fig:traj}
\end{figure}

In this study, we take helium as a prototype system and focus on its ionization within the single active electron model \cite{Tong2005} by a short circularly polarized laser pulse, whose vector potential is given by (atomic units are used unless stated otherwise)
\begin{equation} \label{eq:vector_potential}
	{\bm A}(t) = \frac{A_0}{\sqrt{2}}\cos[4](\frac{\omega t}{2N})\left[\cos(\omega t)\hat{\bm{e}}_x + \sin(\omega t)\hat{\bm{e}}_y\right],
\end{equation}
where $A_0$ is the amplitude of the vector potential corresponding to a peak intensity of $4\times10^{15}$ W/cm$^2$, $\omega=0.045$ a.u.\ is the angular frequency corresponding to a wavelength of 1000 nm, and $N=2$ is the number of optical cycles. The electric field follows as ${\bm F}(t) = -\dot{\bm A}(t)$. For the given laser parameters, the Keldysh parameter is 0.257, and the threshold field strength $F_{\rm th}={I_p^2}/{(4Z)}=0.204$ a.u.\ for OBI to occur is less than the peak field strength $F_0=A_0\omega=0.239$ a.u., indicating ionization in the OBI regime.

To delve into the dynamics of OBI, it is essential to employ the full system Hamiltonian, inclusive of the Coulomb potential. This comprehensive approach can be achieved through an extension of the backpropagation method. In line with the original backpropagation method \cite{Ni2016,Ni2018,Ni2018a,Hofmann2021,Ma2024}, our approach starts with the numerical solution of the time-dependent Schrödinger equation to obtain the ionized wave function. This wave function is subsequently converted into classical trajectories. The final step is the backpropagation of these trajectories along the time axis until specific criteria are fulfilled, as illustrated in Fig.~\ref{fig:traj}. In the TI regime, the velocity criterion $k_\parallel=0$ (velocity in the instantaneous field direction vanishes) is usually used \cite{Ni2018a}, which captures nonadiabatic effects and accurately retrieves tunneling exit information. This has made the backpropagation method a widely used tool in various studies, including the subcycle transfer of linear momentum \cite{Ni2020}, probing of rescattering times \cite{Kim2021}, and the discovery of subcycle conservation laws \cite{Ma2024US}. The method has proven to reliably provide highly differential information of the tunnel exit. However, in the OBI regime, the velocity criterion fails due to high nontunneled fraction \cite{Ni2018}. Consequently, it is necessary to introduce new ways to identify OBI trajectories and new stopping criteria for backpropagation tailored to the OBI regime.

\begin{figure*}[t]
	\centering
	\includegraphics[width=\textwidth]{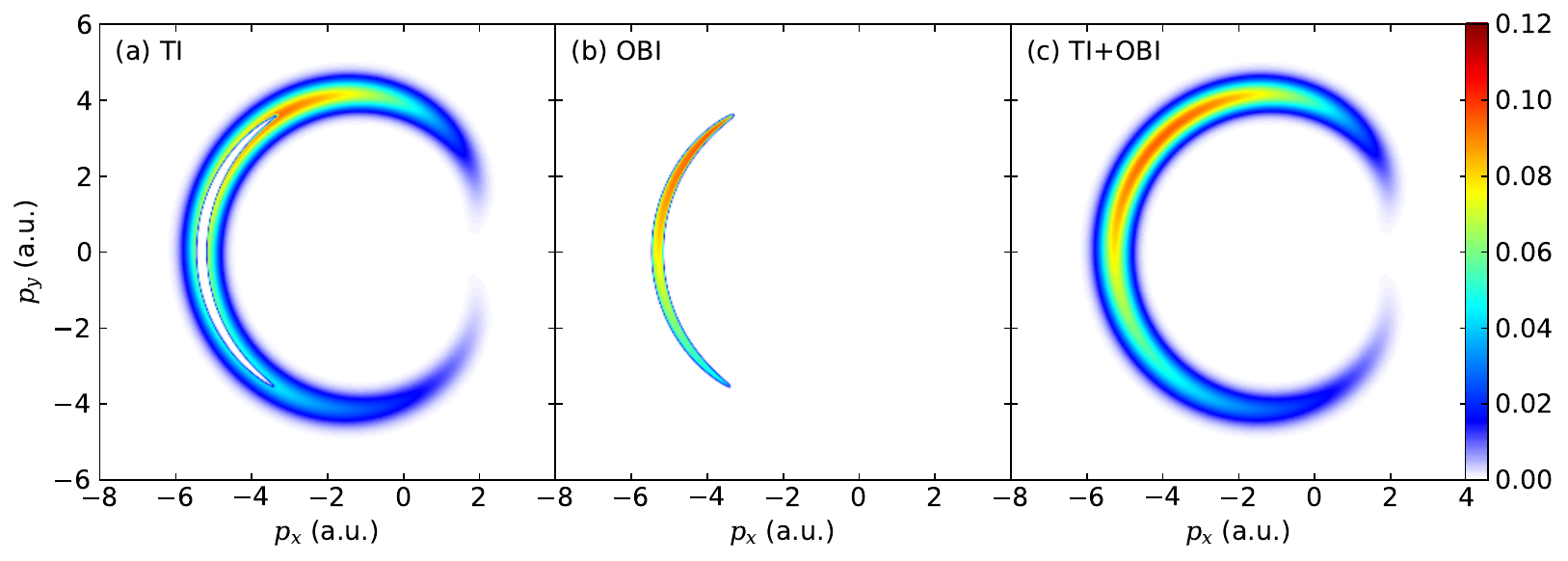}
	\caption{Photoelectron momentum distribution resulting from the ionization of the model helium atom by a two-cycle circularly polarized laser pulse with a wavelength of 1000 nm and a peak intensity of $4.0 \times 10^{15}$ W/cm$^2$. The distributions correspond to: (a) TI, (b) OBI, and (c) the sum of both mechanisms.}
	\label{fig:PMD}
\end{figure*}

To this end, we scrutinize the backpropagating trajectories and identify two distinct categories. The first category includes trajectories where the electron gradually approaches the parent ion during backpropagation and then rebounds from the potential barrier at the tunnel exit, moving away from the parent ion. This represents a typical TI trajectory, with the tunnel exit pinpointed using the velocity criterion $k_\parallel=0$, as depicted in Fig.~\ref{fig:traj}(a). The second category involves trajectories where the electron returns to the proximity of the parent ion during backpropagation and subsequently orbits around it, thereby creating a ``hole'' in the trajectory at the ion, which effectively makes the trajectory ``topological''. This trajectory is characteristic of OBI. The barrier top is identified at the point where the electron speed $k=\sqrt{k_x^2 + k_y^2}$, prior to its orbital motion around the parent ion, attains a local minimum, as shown in Fig.~\ref{fig:traj}(b). This classification based on the trajectory topology allows us to extract dynamic information pertaining to both TI and OBI electrons. Gaining these insights is essential for further analysis, especially regarding their individual contributions to the PMD and the distribution of ionization times.

In Figs.~\ref{fig:PMD}(a) and \ref{fig:PMD}(b), we depict the PMD for TI and OBI electrons, respectively. As expected, the sum of these two distributions constitutes the complete PMD, which is displayed in Fig.~\ref{fig:PMD}(c). This confirms the accuracy of the trajectory classification within the backpropagation method. The use of an ultrashort circularly polarized laser pulse results in an arc-shaped structure in the total PMD, effectively mitigating the impact of intercycle interference effects. Due to the depletion effect under the intense laser field, the ionization probability is biased towards the leading edge of the laser pulse, thereby resulting in the maximum emission occurring along the second quadrant in Fig.~\ref{fig:PMD}(c). We note that backpropagation is applicable even when depletion is present as the same Hamiltonian is used for quantum forward and classical backward propagation. More importantly, the PMD for TI encircles that for OBI. This distinct and orderly separation carries significant physical implications.

To uncover the physical significance behind the PMD for TI and OBI, we introduce a simple conceptual framework. Based on the Ammosov-Delone-Krainov (ADK) theory \cite{Ammosov1986,Ivanov2005}, we understand that photoelectrons acquire an initial transverse momentum $k_\perp$ perpendicular to the field polarization direction at the tunnel exit. Accordingly, one can establish an effective potential barrier model that relates to the electron position $\bm r$ and initial transverse momentum $\bm{k}_\perp$ under a constant electric field $\bm{F}_c$. The energy associated with this effective potential barrier is given by $E = \frac{1}{2}k_\perp^2 - \frac{1}{r} + \bm{r} \cdot \bm{F}_c$, and its distribution exhibits a saddle shape, as depicted in Fig.~\ref{fig:potential}(b). Ignoring nonadiabatic effects \cite{Klaiber2013,Li2016,Eckart2018,Liu2019,Ni2018a}, the electron energy remains constant throughout the tunneling process. OBI occurs for specific values of transverse momentum $k_\perp$ when the ground-state energy exceeds the corresponding saddle surface level. Conversely, the absence of such conditions confines the process to TI exclusively. This model effectively clarifies the PMD characteristics of both TI and OBI electrons. Utilizing the attoclock principle and neglecting the Coulomb potential, the asymptotic momentum of the photoelectron can be expressed as $\bm p = \bm k_\perp -\bm A(t)$. The initial transverse momentum $\bm{k}_\perp$ at the tunnel exit (or barrier top) is thus mirrored into the asymptotic radial momentum, and its presence broadens the asymptotic radial momentum distribution centered near $-\bm A(t)$. Thus, a smaller transverse momentum $k_\perp$, associated with OBI [orange arrow in Fig.~\ref{fig:potential}(b)], results in a PMD concentrated in the central region. In contrast, a larger transverse momentum $k_\perp$, associated with TI [green arrow in Fig.~\ref{fig:potential}(b)], causes the PMD to extend into the peripheral areas. Under conditions of low field strength, OBI does not occur, even when the transverse momentum $k_\perp$ is zero. Consequently, the PMD appears as a large arc-shaped TI momentum spectrum encircling a smaller arc-shaped OBI momentum spectrum. This indicates that the occurrence of OBI, or, the border between TI and OBI, is dependent not only on the electric field strength but also on the initial transverse momentum $k_\perp$ of the photoelectron. Remarkably, this border is sharp. This is because the categorization of TI and OBI is based on trajectory topology, which is a binary condition.

\begin{figure*}[t]
	\centering
	\includegraphics[width=\textwidth]{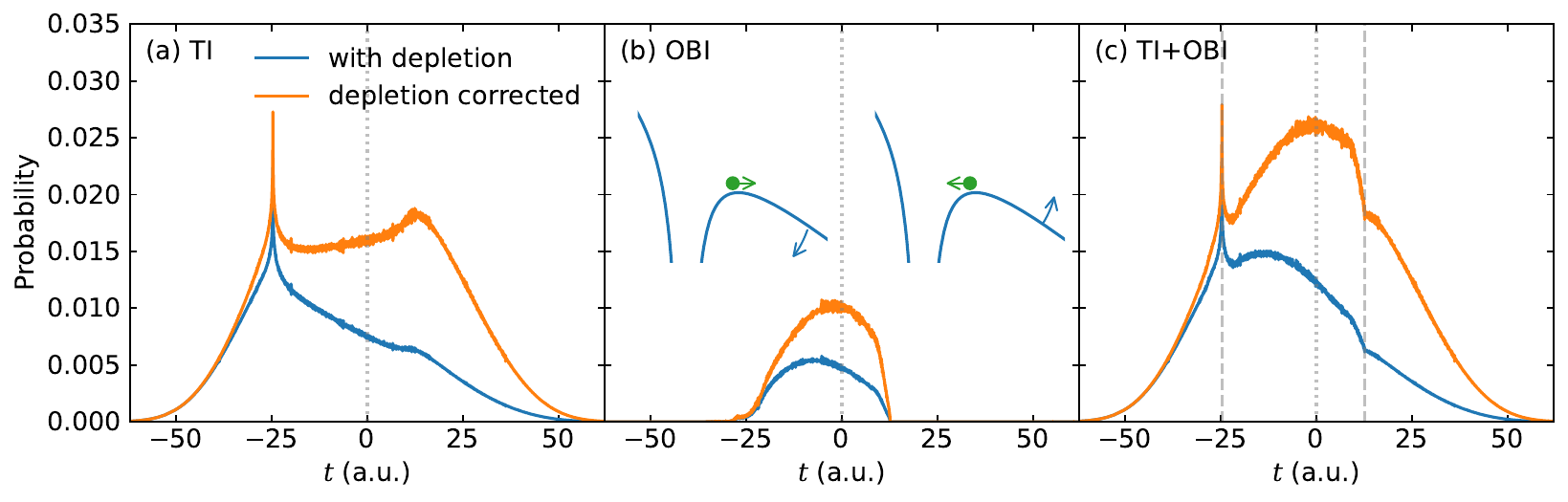}
	\caption{Distribution of ionization times corresponding to (a) TI, (b) OBI, and (c) the sum of both mechanisms. The laser parameters are identical to those used in Fig.~\ref{fig:PMD}. The blue solid lines represent the original results without depletion correction $P(t)$, while the yellow solid lines represent the results with depletion correction $\tilde{P}(t)$. The gray dotted lines indicate the pulse center ($t=0$). The gray dashed lines mark the peak and dip in the ionization time distribution caused by stranded trajectories, delineating the boundary between TI and OBI. The insets in panel (b) provides a schematic explanation for the asymmetric positioning of the peak and dip relative to the laser center.}
	\label{fig:ti_prob}
\end{figure*}

\begin{figure}[b]
	\centering
	\includegraphics[width=\columnwidth]{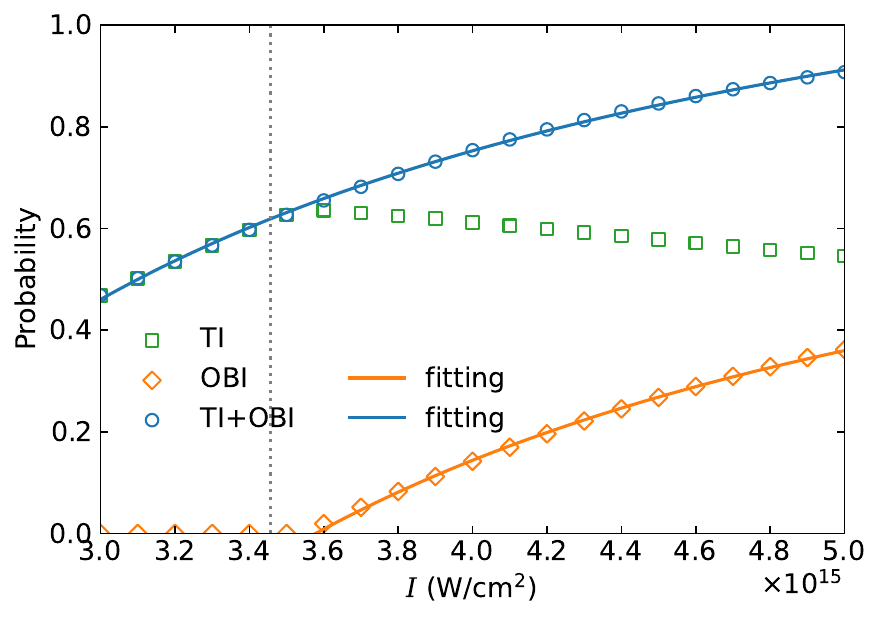}
	\caption{Intensity dependence of the ionization probability for TI, OBI, and the sum of both mechanisms. The orange and blue solid lines represent the fits to the OBI and the sum of TI+OBI data using the function $aI^n + b$, respectively. The gray dotted line indicates the estimated threshold laser intensity, taking into account the Stark shift.}
	\label{fig:I_prob}
\end{figure}

The ionization time distribution for both TI and OBI electrons is discernible using the backpropagation method, as illustrated in Fig.~\ref{fig:ti_prob}. As previously noted, the depletion of the ground state leads to ionization predominantly on the leading edge of the laser field (blue solid lines). To facilitate a more nuanced analysis of the ionization time distribution, we correct for the depletion effect using \cite{Ni2018,Klaiber2022}
\begin{equation} \label{eq:depletion}
	\tilde{P}(t) = \frac{P(t)}{1-\int_{-\infty}^{t}P_{t}(\tau) d\tau},
\end{equation}
where $P_{t}$ is the total ionization rate, and $P$ is the ionization rate associated with the respective mechanisms. After correcting for ground-state depletion, the ionization time distribution (orange solid lines) shows approximate symmetry around the laser peak ($t=0$). This symmetry indicates the absence of significant time delays in both TI and OBI processes. The ionization time for OBI is primarily concentrated near the peak of the laser field where the field strength is high. It is important to note that TI also occurs near the peak of the laser field, with the ionization time distribution for TI displaying a plateau-like structure. This is due to the presence of initial transverse momentum at the tunnel exit, which enables TI to occur even when the electric field strength surpasses the threshold field strength $F_{\rm th}$. Notably, at the boundary between OBI and TI, as seen in Fig.~\ref{fig:ti_prob}(c), the total ionization time distribution exhibits a pronounced peak on the leading edge of the laser field and a significant dip on the trailing edge. This pattern is the consequence of finite propagation times and the fact that at the transition region between OBI and TI a singularity in the ionization time emerges due to trajectories stranded \cite{vandeSand1999,Rodriguez2022} at the top of the effective potential barrier. The observed asymmetric positioning of the peak and dip, or, the border for OBI to occur, relative to the pulse center is a result of the subcycle nonadiabatic effect during the tunneling process \cite{Ni2018a,Li2016}. On the leading edge of the laser field, the potential barrier lowers down as the field strength increases. Electrons initially have an outgoing longitudinal momentum, thus requiring a lower field strength for OBI to occur. Conversely, on the trailing edge of the laser field, the potential barrier rises up as the field strength decreases. Electrons at this point have an incoming longitudinal momentum, necessitating a higher field strength for OBI to take place. This concept is visualized in the insets in Fig.~\ref{fig:ti_prob}(b).

Furthermore, we present the laser intensity dependence of the ionization probability for TI and OBI in Fig.~\ref{fig:I_prob}. At lower light intensities, which are insufficient to induce OBI, TI is the predominant mechanism. As the light intensity exceeds a certain threshold, the probability of OBI increases with rising light intensity. Notably, the probability of TI decreases with increasing light intensity, revealing a competitive interplay between OBI and TI. When the ionization probabilities of the two mechanisms become comparable, their distinct contributions become evident in physical observables, as detailed in our previous analysis. Employing the fitting function $aI^n+b$, we obtain a quantitative characterization of the variations in the probabilities of TI and OBI as functions of light intensity $I$ (in atomic units), expressed as:
\begin{align} \label{eq:Fit}
	P_{\rm OBI} = -0.288I^{-1.758}+0.809,  \\
	P_{\rm TI+OBI} = -0.313I^{-1.377}+1.355.
\end{align}
The light intensity at which $P_{\rm OBI} = 0$ is defined as the threshold light intensity $I_{\rm OBI} = 0.556$ a.u.\ (or $3.577 \times 10^{15}$ W/cm$^2$). For the commonly used threshold electric field $F_{\rm th}$, the corresponding light intensity $I_{\rm th} = 0.454$ a.u.\ (or $2.924 \times 10^{15}$ W/cm$^2$), which is much lower than $I_{\rm OBI}$. To accurately assess the threshold field strength in the OBI regime, we must consider the Stark shift of the ground-state energy. Thus, the form of the threshold field strength becomes
\begin{equation} \label{eq:F_th_obi}
	F_{\rm th}^{\rm Stark}=\frac{\left(I_p^{\rm Stark}\right)^2}{4Z} = \frac{\left[I_p+\frac{1}{2}\alpha \left(F_{\rm th}^{\rm Stark}\right)^2\right]^2}{4Z},
\end{equation}
where $\alpha=1.57$ is the polarizability of the two-dimensional model helium atom, determined numerically by observing the energy shift in response to a range of small external static electric fields \cite{Ni2018a}. Solution of Eq.~\eqref{eq:F_th_obi} gives the threshold field strength $F_{\rm th}^{\rm Stark}=0.222$, and the corresponding threshold laser intensity $I_{\rm th}^{\rm Stark}=0.537$ a.u.\ (or $3.458\times10^{15}$ W/cm$^2$), which is close to $I_{\rm OBI}$ obtained from our numerical calculation. This proximity indicates that the Stark effect is a critical factor that cannot be overlooked in the OBI regime.

In conclusion, our comprehensive investigation into the electron dynamics within the over-barrier ionization (OBI) regime, facilitated by a specially tailored backpropagation method, has yielded significant physical insights. We have illuminated the distinct dynamic behaviors between electrons undergoing OBI and tunneling ionization (TI), allowing for a precise differentiation of their respective contributions to the photoelectron momentum spectrum and ionization time distribution. By leveraging effective potential barriers, we have determined that the occurrence of OBI is dependent not only on the field strength but also on the initial transverse momentum of the photoelectron. Additionally, our study unveiled the dependence of ionization probabilities on light intensity for both TI and OBI, highlighting a competitive interplay. The findings underscore the essential role of Stark shift in determining threshold field strengths. Our study not only enhances our understanding of attosecond science but also paves the way for new frontiers in strong-field physics, especially in the control and manipulation of electron dynamics under extreme conditions.

We would like to thank J.M.\ Rost and U.\ Saalmann for helpful discussions.
This work was supported by the National Natural Science Foundation of China (Grant Nos.\ 92150105, 12474341, 12227807, 12241407, and 12034020), the Science and Technology Commission of Shanghai Municipality (Grant No.\ 23JC1402000), the Shanghai Pilot Program for Basic Research (Grant No.\ TQ20240204), and the Science Foundation of Shandong Province (Grant No.\ 2022HWYQ-073). Numerical computations were in part performed on the ECNU Multifunctional Platform for Innovation (001).

\bibliography{OBI.bib}

\begin{thebibliography}{53}%
\makeatletter
\providecommand \@ifxundefined [1]{%
 \@ifx{#1\undefined}
}%
\providecommand \@ifnum [1]{%
 \ifnum #1\expandafter \@firstoftwo
 \else \expandafter \@secondoftwo
 \fi
}%
\providecommand \@ifx [1]{%
 \ifx #1\expandafter \@firstoftwo
 \else \expandafter \@secondoftwo
 \fi
}%
\providecommand \natexlab [1]{#1}%
\providecommand \enquote  [1]{``#1''}%
\providecommand \bibnamefont  [1]{#1}%
\providecommand \bibfnamefont [1]{#1}%
\providecommand \citenamefont [1]{#1}%
\providecommand \href@noop [0]{\@secondoftwo}%
\providecommand \href [0]{\begingroup \@sanitize@url \@href}%
\providecommand \@href[1]{\@@startlink{#1}\@@href}%
\providecommand \@@href[1]{\endgroup#1\@@endlink}%
\providecommand \@sanitize@url [0]{\catcode `\\12\catcode `\$12\catcode
  `\&12\catcode `\#12\catcode `\^12\catcode `\_12\catcode `\%12\relax}%
\providecommand \@@startlink[1]{}%
\providecommand \@@endlink[0]{}%
\providecommand \url  [0]{\begingroup\@sanitize@url \@url }%
\providecommand \@url [1]{\endgroup\@href {#1}{\urlprefix }}%
\providecommand \urlprefix  [0]{URL }%
\providecommand \Eprint [0]{\href }%
\providecommand \doibase [0]{https://doi.org/}%
\providecommand \selectlanguage [0]{\@gobble}%
\providecommand \bibinfo  [0]{\@secondoftwo}%
\providecommand \bibfield  [0]{\@secondoftwo}%
\providecommand \translation [1]{[#1]}%
\providecommand \BibitemOpen [0]{}%
\providecommand \bibitemStop [0]{}%
\providecommand \bibitemNoStop [0]{.\EOS\space}%
\providecommand \EOS [0]{\spacefactor3000\relax}%
\providecommand \BibitemShut  [1]{\csname bibitem#1\endcsname}%
\let\auto@bib@innerbib\@empty
%</preamble>
\bibitem [{\citenamefont {Agostini}\ \emph {et~al.}(1979)\citenamefont
  {Agostini}, \citenamefont {Fabre}, \citenamefont {Mainfray}, \citenamefont
  {Petite},\ and\ \citenamefont {Rahman}}]{Agostini1979}%
  \BibitemOpen
  \bibfield  {author} {\bibinfo {author} {\bibfnamefont {P.}~\bibnamefont
  {Agostini}}, \bibinfo {author} {\bibfnamefont {F.}~\bibnamefont {Fabre}},
  \bibinfo {author} {\bibfnamefont {G.}~\bibnamefont {Mainfray}}, \bibinfo
  {author} {\bibfnamefont {G.}~\bibnamefont {Petite}},\ and\ \bibinfo {author}
  {\bibfnamefont {N.~K.}\ \bibnamefont {Rahman}},\ }\bibfield  {title}
  {\bibinfo {title} {Free-free transitions following six-photon ionization of
  xenon atoms},\ }\href {https://doi.org/10.1103/PhysRevLett.42.1127}
  {\bibfield  {journal} {\bibinfo  {journal} {Phys. Rev. Lett.}\ }\textbf
  {\bibinfo {volume} {42}},\ \bibinfo {pages} {1127} (\bibinfo {year}
  {1979})}\BibitemShut {NoStop}%
\bibitem [{\citenamefont {Eberly}\ \emph {et~al.}(1991)\citenamefont {Eberly},
  \citenamefont {Javanainen},\ and\ \citenamefont
  {Rza{\.z}ewski}}]{Eberly1991}%
  \BibitemOpen
  \bibfield  {author} {\bibinfo {author} {\bibfnamefont {J.~H.}\ \bibnamefont
  {Eberly}}, \bibinfo {author} {\bibfnamefont {J.}~\bibnamefont {Javanainen}},\
  and\ \bibinfo {author} {\bibfnamefont {K.}~\bibnamefont {Rza{\.z}ewski}},\
  }\bibfield  {title} {\bibinfo {title} {Above-threshold ionization},\ }\href
  {https://doi.org/https://doi.org/10.1016/0370-1573(91)90131-5} {\bibfield
  {journal} {\bibinfo  {journal} {Phys. Rep.}\ }\textbf {\bibinfo {volume}
  {204}},\ \bibinfo {pages} {331} (\bibinfo {year} {1991})}\BibitemShut
  {NoStop}%
\bibitem [{\citenamefont {Becker}\ \emph {et~al.}(2002)\citenamefont {Becker},
  \citenamefont {Grasbon}, \citenamefont {Kopold}, \citenamefont {Milošević},
  \citenamefont {Paulus},\ and\ \citenamefont {Walther}}]{Becker2002}%
  \BibitemOpen
  \bibfield  {author} {\bibinfo {author} {\bibfnamefont {W.}~\bibnamefont
  {Becker}}, \bibinfo {author} {\bibfnamefont {F.}~\bibnamefont {Grasbon}},
  \bibinfo {author} {\bibfnamefont {R.}~\bibnamefont {Kopold}}, \bibinfo
  {author} {\bibfnamefont {D.}~\bibnamefont {Milošević}}, \bibinfo {author}
  {\bibfnamefont {G.}~\bibnamefont {Paulus}},\ and\ \bibinfo {author}
  {\bibfnamefont {H.}~\bibnamefont {Walther}},\ }\bibfield  {title} {\bibinfo
  {title} {Above-threshold ionization: From classical features to quantum
  effects},\ }\href
  {https://doi.org/https://doi.org/10.1016/S1049-250X(02)80006-4} {\bibfield
  {journal} {\bibinfo  {journal} {Adv. At. Mol. Opt. Phys.}\ }\textbf {\bibinfo
  {volume} {48}},\ \bibinfo {pages} {35} (\bibinfo {year} {2002})}\BibitemShut
  {NoStop}%
\bibitem [{\citenamefont {Fittinghoff}\ \emph {et~al.}(1992)\citenamefont
  {Fittinghoff}, \citenamefont {Bolton}, \citenamefont {Chang},\ and\
  \citenamefont {Kulander}}]{Fittinghoff1992}%
  \BibitemOpen
  \bibfield  {author} {\bibinfo {author} {\bibfnamefont {D.~N.}\ \bibnamefont
  {Fittinghoff}}, \bibinfo {author} {\bibfnamefont {P.~R.}\ \bibnamefont
  {Bolton}}, \bibinfo {author} {\bibfnamefont {B.}~\bibnamefont {Chang}},\ and\
  \bibinfo {author} {\bibfnamefont {K.~C.}\ \bibnamefont {Kulander}},\
  }\bibfield  {title} {\bibinfo {title} {Observation of nonsequential double
  ionization of helium with optical tunneling},\ }\href
  {https://doi.org/10.1103/PhysRevLett.69.2642} {\bibfield  {journal} {\bibinfo
   {journal} {Phys. Rev. Lett.}\ }\textbf {\bibinfo {volume} {69}},\ \bibinfo
  {pages} {2642} (\bibinfo {year} {1992})}\BibitemShut {NoStop}%
\bibitem [{\citenamefont {Walker}\ \emph {et~al.}(1994)\citenamefont {Walker},
  \citenamefont {Sheehy}, \citenamefont {DiMauro}, \citenamefont {Agostini},
  \citenamefont {Schafer},\ and\ \citenamefont {Kulander}}]{Walker1994}%
  \BibitemOpen
  \bibfield  {author} {\bibinfo {author} {\bibfnamefont {B.}~\bibnamefont
  {Walker}}, \bibinfo {author} {\bibfnamefont {B.}~\bibnamefont {Sheehy}},
  \bibinfo {author} {\bibfnamefont {L.~F.}\ \bibnamefont {DiMauro}}, \bibinfo
  {author} {\bibfnamefont {P.}~\bibnamefont {Agostini}}, \bibinfo {author}
  {\bibfnamefont {K.~J.}\ \bibnamefont {Schafer}},\ and\ \bibinfo {author}
  {\bibfnamefont {K.~C.}\ \bibnamefont {Kulander}},\ }\bibfield  {title}
  {\bibinfo {title} {Precision measurement of strong field double ionization of
  helium},\ }\href {https://doi.org/10.1103/PhysRevLett.73.1227} {\bibfield
  {journal} {\bibinfo  {journal} {Phys. Rev. Lett.}\ }\textbf {\bibinfo
  {volume} {73}},\ \bibinfo {pages} {1227} (\bibinfo {year}
  {1994})}\BibitemShut {NoStop}%
\bibitem [{\citenamefont {Ferray}\ \emph {et~al.}(1988)\citenamefont {Ferray},
  \citenamefont {L'Huillier}, \citenamefont {Li}, \citenamefont {Lompre},
  \citenamefont {Mainfray},\ and\ \citenamefont {Manus}}]{Ferray1988}%
  \BibitemOpen
  \bibfield  {author} {\bibinfo {author} {\bibfnamefont {M.}~\bibnamefont
  {Ferray}}, \bibinfo {author} {\bibfnamefont {A.}~\bibnamefont {L'Huillier}},
  \bibinfo {author} {\bibfnamefont {X.~F.}\ \bibnamefont {Li}}, \bibinfo
  {author} {\bibfnamefont {L.~A.}\ \bibnamefont {Lompre}}, \bibinfo {author}
  {\bibfnamefont {G.}~\bibnamefont {Mainfray}},\ and\ \bibinfo {author}
  {\bibfnamefont {C.}~\bibnamefont {Manus}},\ }\bibfield  {title} {\bibinfo
  {title} {Multiple-harmonic conversion of 1064 nm radiation in rare gases},\
  }\href {https://doi.org/10.1088/0953-4075/21/3/001} {\bibfield  {journal}
  {\bibinfo  {journal} {J. Phys. B}\ }\textbf {\bibinfo {volume} {21}},\
  \bibinfo {pages} {L31} (\bibinfo {year} {1988})}\BibitemShut {NoStop}%
\bibitem [{\citenamefont {Krause}\ \emph {et~al.}(1992)\citenamefont {Krause},
  \citenamefont {Schafer},\ and\ \citenamefont {Kulander}}]{Krause1992}%
  \BibitemOpen
  \bibfield  {author} {\bibinfo {author} {\bibfnamefont {J.~L.}\ \bibnamefont
  {Krause}}, \bibinfo {author} {\bibfnamefont {K.~J.}\ \bibnamefont
  {Schafer}},\ and\ \bibinfo {author} {\bibfnamefont {K.~C.}\ \bibnamefont
  {Kulander}},\ }\bibfield  {title} {\bibinfo {title} {High-order harmonic
  generation from atoms and ions in the high intensity regime},\ }\href
  {https://doi.org/10.1103/PhysRevLett.68.3535} {\bibfield  {journal} {\bibinfo
   {journal} {Phys. Rev. Lett.}\ }\textbf {\bibinfo {volume} {68}},\ \bibinfo
  {pages} {3535} (\bibinfo {year} {1992})}\BibitemShut {NoStop}%
\bibitem [{\citenamefont {Macklin}\ \emph {et~al.}(1993)\citenamefont
  {Macklin}, \citenamefont {Kmetec},\ and\ \citenamefont
  {Gordon}}]{Macklin1993}%
  \BibitemOpen
  \bibfield  {author} {\bibinfo {author} {\bibfnamefont {J.~J.}\ \bibnamefont
  {Macklin}}, \bibinfo {author} {\bibfnamefont {J.~D.}\ \bibnamefont
  {Kmetec}},\ and\ \bibinfo {author} {\bibfnamefont {C.~L.}\ \bibnamefont
  {Gordon}},\ }\bibfield  {title} {\bibinfo {title} {High-order harmonic
  generation using intense femtosecond pulses},\ }\href
  {https://doi.org/10.1103/PhysRevLett.70.766} {\bibfield  {journal} {\bibinfo
  {journal} {Phys. Rev. Lett.}\ }\textbf {\bibinfo {volume} {70}},\ \bibinfo
  {pages} {766} (\bibinfo {year} {1993})}\BibitemShut {NoStop}%
\bibitem [{\citenamefont {Corkum}(1993)}]{Corkum1993}%
  \BibitemOpen
  \bibfield  {author} {\bibinfo {author} {\bibfnamefont {P.~B.}\ \bibnamefont
  {Corkum}},\ }\bibfield  {title} {\bibinfo {title} {Plasma perspective on
  strong field multiphoton ionization},\ }\href
  {https://doi.org/10.1103/PhysRevLett.71.1994} {\bibfield  {journal} {\bibinfo
   {journal} {Phys. Rev. Lett.}\ }\textbf {\bibinfo {volume} {71}},\ \bibinfo
  {pages} {1994} (\bibinfo {year} {1993})}\BibitemShut {NoStop}%
\bibitem [{\citenamefont {Lewenstein}\ \emph {et~al.}(1994)\citenamefont
  {Lewenstein}, \citenamefont {Balcou}, \citenamefont {Ivanov}, \citenamefont
  {L'Huillier},\ and\ \citenamefont {Corkum}}]{Lewenstein1994}%
  \BibitemOpen
  \bibfield  {author} {\bibinfo {author} {\bibfnamefont {M.}~\bibnamefont
  {Lewenstein}}, \bibinfo {author} {\bibfnamefont {{\relax Ph}.}~\bibnamefont
  {Balcou}}, \bibinfo {author} {\bibfnamefont {M.~{\relax Yu}.}\ \bibnamefont
  {Ivanov}}, \bibinfo {author} {\bibfnamefont {A.}~\bibnamefont {L'Huillier}},\
  and\ \bibinfo {author} {\bibfnamefont {P.~B.}\ \bibnamefont {Corkum}},\
  }\bibfield  {title} {\bibinfo {title} {Theory of high-harmonic generation by
  low-frequency laser fields},\ }\href
  {https://doi.org/10.1103/PhysRevA.49.2117} {\bibfield  {journal} {\bibinfo
  {journal} {Phys. Rev. A}\ }\textbf {\bibinfo {volume} {49}},\ \bibinfo
  {pages} {2117} (\bibinfo {year} {1994})}\BibitemShut {NoStop}%
\bibitem [{\citenamefont {Popmintchev}\ \emph {et~al.}(2010)\citenamefont
  {Popmintchev}, \citenamefont {Chen}, \citenamefont {Arpin}, \citenamefont
  {Murnane},\ and\ \citenamefont {Kapteyn}}]{Popmintchev2010}%
  \BibitemOpen
  \bibfield  {author} {\bibinfo {author} {\bibfnamefont {T.}~\bibnamefont
  {Popmintchev}}, \bibinfo {author} {\bibfnamefont {M.-C.}\ \bibnamefont
  {Chen}}, \bibinfo {author} {\bibfnamefont {P.}~\bibnamefont {Arpin}},
  \bibinfo {author} {\bibfnamefont {M.~M.}\ \bibnamefont {Murnane}},\ and\
  \bibinfo {author} {\bibfnamefont {H.~C.}\ \bibnamefont {Kapteyn}},\
  }\bibfield  {title} {\bibinfo {title} {The attosecond nonlinear optics of
  bright coherent {{X-ray}} generation},\ }\href
  {https://doi.org/10.1038/nphoton.2010.256} {\bibfield  {journal} {\bibinfo
  {journal} {Nat. Photonics}\ }\textbf {\bibinfo {volume} {4}},\ \bibinfo
  {pages} {822} (\bibinfo {year} {2010})}\BibitemShut {NoStop}%
\bibitem [{\citenamefont {Keldysh}(1965)}]{Keldysh1965}%
  \BibitemOpen
  \bibfield  {author} {\bibinfo {author} {\bibfnamefont {L.~V.}\ \bibnamefont
  {Keldysh}},\ }\bibfield  {title} {\bibinfo {title} {Ionization in the field
  of a strong electromagnetic wave},\ }\href@noop {} {\bibfield  {journal}
  {\bibinfo  {journal} {Sov. Phys. JETP}\ }\textbf {\bibinfo {volume} {20}},\
  \bibinfo {pages} {1307} (\bibinfo {year} {1965})}\BibitemShut {NoStop}%
\bibitem [{\citenamefont {Ammosov}\ \emph {et~al.}(1986)\citenamefont
  {Ammosov}, \citenamefont {Delone},\ and\ \citenamefont
  {Krainov}}]{Ammosov1986}%
  \BibitemOpen
  \bibfield  {author} {\bibinfo {author} {\bibfnamefont {M.~V.}\ \bibnamefont
  {Ammosov}}, \bibinfo {author} {\bibfnamefont {N.~B.}\ \bibnamefont
  {Delone}},\ and\ \bibinfo {author} {\bibfnamefont {V.~P.}\ \bibnamefont
  {Krainov}},\ }\bibfield  {title} {\bibinfo {title} {Tunnel ionization of
  complex atoms and of atomic ions in an alternating electromagnetic field},\
  }\href@noop {} {\bibfield  {journal} {\bibinfo  {journal} {Sov. Phys. JETP}\
  }\textbf {\bibinfo {volume} {64}},\ \bibinfo {pages} {1191} (\bibinfo {year}
  {1986})}\BibitemShut {NoStop}%
\bibitem [{\citenamefont {Ivanov}\ \emph {et~al.}(2005)\citenamefont {Ivanov},
  \citenamefont {Spanner},\ and\ \citenamefont {Smirnova}}]{Ivanov2005}%
  \BibitemOpen
  \bibfield  {author} {\bibinfo {author} {\bibfnamefont {M.~Y.}\ \bibnamefont
  {Ivanov}}, \bibinfo {author} {\bibfnamefont {M.}~\bibnamefont {Spanner}},\
  and\ \bibinfo {author} {\bibfnamefont {O.}~\bibnamefont {Smirnova}},\
  }\bibfield  {title} {\bibinfo {title} {Anatomy of strong field ionization},\
  }\href {https://doi.org/10.1080/0950034042000275360} {\bibfield  {journal}
  {\bibinfo  {journal} {J. Mod. Opt.}\ }\textbf {\bibinfo {volume} {52}},\
  \bibinfo {pages} {165} (\bibinfo {year} {2005})}\BibitemShut {NoStop}%
\bibitem [{\citenamefont {Popov}(2004)}]{Popov2004}%
  \BibitemOpen
  \bibfield  {author} {\bibinfo {author} {\bibfnamefont {V.~S.}\ \bibnamefont
  {Popov}},\ }\bibfield  {title} {\bibinfo {title} {Tunnel and multiphoton
  ionization of atoms and ions in a strong laser field (keldysh theory)},\
  }\href {https://doi.org/10.1070/PU2004v047n09ABEH001812} {\bibfield
  {journal} {\bibinfo  {journal} {Phys.-Usp.}\ }\textbf {\bibinfo {volume}
  {47}},\ \bibinfo {pages} {855} (\bibinfo {year} {2004})}\BibitemShut
  {NoStop}%
\bibitem [{\citenamefont {Popruzhenko}(2014)}]{Popruzhenko2014}%
  \BibitemOpen
  \bibfield  {author} {\bibinfo {author} {\bibfnamefont {S.~V.}\ \bibnamefont
  {Popruzhenko}},\ }\bibfield  {title} {\bibinfo {title} {Keldysh theory of
  strong field ionization: history, applications, difficulties and
  perspectives},\ }\href {https://doi.org/10.1088/0953-4075/47/20/204001}
  {\bibfield  {journal} {\bibinfo  {journal} {J. Phys. B.}\ }\textbf {\bibinfo
  {volume} {47}},\ \bibinfo {pages} {204001} (\bibinfo {year}
  {2014})}\BibitemShut {NoStop}%
\bibitem [{\citenamefont {Keldysh}(2017)}]{Keldysh2017}%
  \BibitemOpen
  \bibfield  {author} {\bibinfo {author} {\bibfnamefont {L.~V.}\ \bibnamefont
  {Keldysh}},\ }\bibfield  {title} {\bibinfo {title} {Multiphoton ionization by
  a very short pulse},\ }\href {https://doi.org/10.3367/UFNe.2017.10.038229}
  {\bibfield  {journal} {\bibinfo  {journal} {Phys.-Usp.}\ }\textbf {\bibinfo
  {volume} {60}},\ \bibinfo {pages} {1187} (\bibinfo {year}
  {2017})}\BibitemShut {NoStop}%
\bibitem [{\citenamefont {Ma}\ \emph {et~al.}(2024{\natexlab{a}})\citenamefont
  {Ma}, \citenamefont {Ni},\ and\ \citenamefont {Wu}}]{Ma2024}%
  \BibitemOpen
  \bibfield  {author} {\bibinfo {author} {\bibfnamefont {Y.}~\bibnamefont
  {Ma}}, \bibinfo {author} {\bibfnamefont {H.}~\bibnamefont {Ni}},\ and\
  \bibinfo {author} {\bibfnamefont {J.}~\bibnamefont {Wu}},\ }\bibfield
  {title} {\bibinfo {title} {{Attosecond Ionization Time Delays in Strong-Field
  Physics}},\ }\href {https://doi.org/10.1088/1674-1056/ad0e5d} {\bibfield
  {journal} {\bibinfo  {journal} {Chin. Phys. B}\ }\textbf {\bibinfo {volume}
  {33}},\ \bibinfo {pages} {013201} (\bibinfo {year}
  {2024}{\natexlab{a}})}\BibitemShut {NoStop}%
\bibitem [{\citenamefont {Augst}\ \emph {et~al.}(1989)\citenamefont {Augst},
  \citenamefont {Strickland}, \citenamefont {Meyerhofer}, \citenamefont
  {Chin},\ and\ \citenamefont {Eberly}}]{Augst1989}%
  \BibitemOpen
  \bibfield  {author} {\bibinfo {author} {\bibfnamefont {S.}~\bibnamefont
  {Augst}}, \bibinfo {author} {\bibfnamefont {D.}~\bibnamefont {Strickland}},
  \bibinfo {author} {\bibfnamefont {D.~D.}\ \bibnamefont {Meyerhofer}},
  \bibinfo {author} {\bibfnamefont {S.~L.}\ \bibnamefont {Chin}},\ and\
  \bibinfo {author} {\bibfnamefont {J.~H.}\ \bibnamefont {Eberly}},\ }\bibfield
   {title} {\bibinfo {title} {Tunneling ionization of noble gases in a
  high-intensity laser field},\ }\href
  {https://doi.org/10.1103/PhysRevLett.63.2212} {\bibfield  {journal} {\bibinfo
   {journal} {Phys. Rev. Lett.}\ }\textbf {\bibinfo {volume} {63}},\ \bibinfo
  {pages} {2212} (\bibinfo {year} {1989})}\BibitemShut {NoStop}%
\bibitem [{\citenamefont {Augst}\ \emph {et~al.}(1991)\citenamefont {Augst},
  \citenamefont {Meyerhofer}, \citenamefont {Strickland},\ and\ \citenamefont
  {Chin}}]{Augst1991}%
  \BibitemOpen
  \bibfield  {author} {\bibinfo {author} {\bibfnamefont {S.}~\bibnamefont
  {Augst}}, \bibinfo {author} {\bibfnamefont {D.~D.}\ \bibnamefont
  {Meyerhofer}}, \bibinfo {author} {\bibfnamefont {D.}~\bibnamefont
  {Strickland}},\ and\ \bibinfo {author} {\bibfnamefont {S.~L.}\ \bibnamefont
  {Chin}},\ }\bibfield  {title} {\bibinfo {title} {Laser ionization of noble
  gases by coulomb-barrier suppression},\ }\href
  {https://doi.org/10.1364/JOSAB.8.000858} {\bibfield  {journal} {\bibinfo
  {journal} {J. Opt. Soc. Am. B}\ }\textbf {\bibinfo {volume} {8}},\ \bibinfo
  {pages} {858} (\bibinfo {year} {1991})}\BibitemShut {NoStop}%
\bibitem [{\citenamefont {Auguste}\ \emph {et~al.}(1992)\citenamefont
  {Auguste}, \citenamefont {Monot}, \citenamefont {Lompre}, \citenamefont
  {Mainfray},\ and\ \citenamefont {Manus}}]{Auguste1992}%
  \BibitemOpen
  \bibfield  {author} {\bibinfo {author} {\bibfnamefont {T.}~\bibnamefont
  {Auguste}}, \bibinfo {author} {\bibfnamefont {P.}~\bibnamefont {Monot}},
  \bibinfo {author} {\bibfnamefont {L.~A.}\ \bibnamefont {Lompre}}, \bibinfo
  {author} {\bibfnamefont {G.}~\bibnamefont {Mainfray}},\ and\ \bibinfo
  {author} {\bibfnamefont {C.}~\bibnamefont {Manus}},\ }\bibfield  {title}
  {\bibinfo {title} {Multiply charged ions produced in noble gases by a 1 ps
  laser pulse at lambda =1053 nm},\ }\href
  {https://doi.org/10.1088/0953-4075/25/20/015} {\bibfield  {journal} {\bibinfo
   {journal} {J. Phys. B.}\ }\textbf {\bibinfo {volume} {25}},\ \bibinfo
  {pages} {4181} (\bibinfo {year} {1992})}\BibitemShut {NoStop}%
\bibitem [{\citenamefont {Bauer}(1997)}]{Bauer1997}%
  \BibitemOpen
  \bibfield  {author} {\bibinfo {author} {\bibfnamefont {D.}~\bibnamefont
  {Bauer}},\ }\bibfield  {title} {\bibinfo {title} {Ejection energy of
  photoelectrons in strong-field ionization},\ }\href
  {https://doi.org/10.1103/PhysRevA.55.2180} {\bibfield  {journal} {\bibinfo
  {journal} {Phys. Rev. A}\ }\textbf {\bibinfo {volume} {55}},\ \bibinfo
  {pages} {2180} (\bibinfo {year} {1997})}\BibitemShut {NoStop}%
\bibitem [{\citenamefont {G{\"o}rlinger}\ \emph {et~al.}(2000)\citenamefont
  {G{\"o}rlinger}, \citenamefont {Plagne},\ and\ \citenamefont
  {Kull}}]{Gorlinger2000}%
  \BibitemOpen
  \bibfield  {author} {\bibinfo {author} {\bibfnamefont {J.}~\bibnamefont
  {G{\"o}rlinger}}, \bibinfo {author} {\bibfnamefont {L.}~\bibnamefont
  {Plagne}},\ and\ \bibinfo {author} {\bibfnamefont {H.-J.}\ \bibnamefont
  {Kull}},\ }\bibfield  {title} {\bibinfo {title} {Above-barrier ionization and
  quantum interference in strong laser fields},\ }\href
  {https://doi.org/https://doi.org/10.1007/s003400000345} {\bibfield  {journal}
  {\bibinfo  {journal} {Appl. Phys. B}\ }\textbf {\bibinfo {volume} {71}},\
  \bibinfo {pages} {331} (\bibinfo {year} {2000})}\BibitemShut {NoStop}%
\bibitem [{\citenamefont {Krainov}\ and\ \citenamefont
  {Shokri}(1995)}]{Krainov1995}%
  \BibitemOpen
  \bibfield  {author} {\bibinfo {author} {\bibfnamefont {V.}~\bibnamefont
  {Krainov}}\ and\ \bibinfo {author} {\bibfnamefont {B.}~\bibnamefont
  {Shokri}},\ }\bibfield  {title} {\bibinfo {title} {Energy and angular
  distributions of electrons resulting from barrier-suppression ionization of
  atoms by strong low-frequency radiation},\ }\href@noop {} {\bibfield
  {journal} {\bibinfo  {journal} {J. Exp. Theor. Phys.}\ }\textbf {\bibinfo
  {volume} {80}},\ \bibinfo {pages} {657} (\bibinfo {year} {1995})}\BibitemShut
  {NoStop}%
\bibitem [{\citenamefont {Krainov}(1997)}]{Krainov1997}%
  \BibitemOpen
  \bibfield  {author} {\bibinfo {author} {\bibfnamefont {V.~P.}\ \bibnamefont
  {Krainov}},\ }\bibfield  {title} {\bibinfo {title} {Ionization rates and
  energy and angular distributions at the barrier-suppression ionization of
  complex atoms and atomic ions},\ }\href
  {https://doi.org/10.1364/JOSAB.14.000425} {\bibfield  {journal} {\bibinfo
  {journal} {J. Opt. Soc. Am. B}\ }\textbf {\bibinfo {volume} {14}},\ \bibinfo
  {pages} {425} (\bibinfo {year} {1997})}\BibitemShut {NoStop}%
\bibitem [{\citenamefont {Delone}\ and\ \citenamefont
  {Krainov}(1998)}]{Delone1998}%
  \BibitemOpen
  \bibfield  {author} {\bibinfo {author} {\bibfnamefont {N.~B.}\ \bibnamefont
  {Delone}}\ and\ \bibinfo {author} {\bibfnamefont {V.~P.}\ \bibnamefont
  {Krainov}},\ }\bibfield  {title} {\bibinfo {title} {Tunneling and
  barrier-suppression ionization of atoms and ions in a laser radiation
  field},\ }\href {https://doi.org/10.1070/PU1998v041n05ABEH000393} {\bibfield
  {journal} {\bibinfo  {journal} {Phys.-Usp.}\ }\textbf {\bibinfo {volume}
  {41}},\ \bibinfo {pages} {469} (\bibinfo {year} {1998})}\BibitemShut
  {NoStop}%
\bibitem [{\citenamefont {Klaiber}\ \emph {et~al.}(2024)\citenamefont
  {Klaiber}, \citenamefont {Hatsagortsyan},\ and\ \citenamefont
  {Keitel}}]{Klaiber2024}%
  \BibitemOpen
  \bibfield  {author} {\bibinfo {author} {\bibfnamefont {M.}~\bibnamefont
  {Klaiber}}, \bibinfo {author} {\bibfnamefont {K.~Z.}\ \bibnamefont
  {Hatsagortsyan}},\ and\ \bibinfo {author} {\bibfnamefont {C.~H.}\
  \bibnamefont {Keitel}},\ }\bibfield  {title} {\bibinfo {title} {Generalized
  analytical description of relativistic strong-field ionization},\ }\href
  {https://doi.org/10.1103/PhysRevA.110.023103} {\bibfield  {journal} {\bibinfo
   {journal} {Phys. Rev. A}\ }\textbf {\bibinfo {volume} {110}},\ \bibinfo
  {pages} {023103} (\bibinfo {year} {2024})}\BibitemShut {NoStop}%
\bibitem [{\citenamefont {Bauer}\ and\ \citenamefont
  {Mulser}(1999)}]{Bauer1999}%
  \BibitemOpen
  \bibfield  {author} {\bibinfo {author} {\bibfnamefont {D.}~\bibnamefont
  {Bauer}}\ and\ \bibinfo {author} {\bibfnamefont {P.}~\bibnamefont {Mulser}},\
  }\bibfield  {title} {\bibinfo {title} {Exact field ionization rates in the
  barrier-suppression regime from numerical time-dependent
  schr\"odinger-equation calculations},\ }\href
  {https://doi.org/10.1103/PhysRevA.59.569} {\bibfield  {journal} {\bibinfo
  {journal} {Phys. Rev. A}\ }\textbf {\bibinfo {volume} {59}},\ \bibinfo
  {pages} {569} (\bibinfo {year} {1999})}\BibitemShut {NoStop}%
\bibitem [{\citenamefont {Scrinzi}\ \emph {et~al.}(1999)\citenamefont
  {Scrinzi}, \citenamefont {Geissler},\ and\ \citenamefont
  {Brabec}}]{Scrinzi1999}%
  \BibitemOpen
  \bibfield  {author} {\bibinfo {author} {\bibfnamefont {A.}~\bibnamefont
  {Scrinzi}}, \bibinfo {author} {\bibfnamefont {M.}~\bibnamefont {Geissler}},\
  and\ \bibinfo {author} {\bibfnamefont {T.}~\bibnamefont {Brabec}},\
  }\bibfield  {title} {\bibinfo {title} {Ionization above the coulomb
  barrier},\ }\href {https://doi.org/10.1103/PhysRevLett.83.706} {\bibfield
  {journal} {\bibinfo  {journal} {Phys. Rev. Lett.}\ }\textbf {\bibinfo
  {volume} {83}},\ \bibinfo {pages} {706} (\bibinfo {year} {1999})}\BibitemShut
  {NoStop}%
\bibitem [{\citenamefont {Tong}\ and\ \citenamefont {Lin}(2005)}]{Tong2005}%
  \BibitemOpen
  \bibfield  {author} {\bibinfo {author} {\bibfnamefont {X.~M.}\ \bibnamefont
  {Tong}}\ and\ \bibinfo {author} {\bibfnamefont {C.~D.}\ \bibnamefont {Lin}},\
  }\bibfield  {title} {\bibinfo {title} {Empirical formula for static field
  ionization rates of atoms and molecules by lasers in the barrier-suppression
  regime},\ }\href {https://doi.org/10.1088/0953-4075/38/15/001} {\bibfield
  {journal} {\bibinfo  {journal} {J. Phys. B.}\ }\textbf {\bibinfo {volume}
  {38}},\ \bibinfo {pages} {2593} (\bibinfo {year} {2005})}\BibitemShut
  {NoStop}%
\bibitem [{\citenamefont {Zhang}\ \emph {et~al.}(2014)\citenamefont {Zhang},
  \citenamefont {Lan},\ and\ \citenamefont {Lu}}]{Zhang2014}%
  \BibitemOpen
  \bibfield  {author} {\bibinfo {author} {\bibfnamefont {Q.}~\bibnamefont
  {Zhang}}, \bibinfo {author} {\bibfnamefont {P.}~\bibnamefont {Lan}},\ and\
  \bibinfo {author} {\bibfnamefont {P.}~\bibnamefont {Lu}},\ }\bibfield
  {title} {\bibinfo {title} {Empirical formula for over-barrier strong-field
  ionization},\ }\href {https://doi.org/10.1103/PhysRevA.90.043410} {\bibfield
  {journal} {\bibinfo  {journal} {Phys. Rev. A}\ }\textbf {\bibinfo {volume}
  {90}},\ \bibinfo {pages} {043410} (\bibinfo {year} {2014})}\BibitemShut
  {NoStop}%
\bibitem [{\citenamefont {Bauer}(2017)}]{Bauer2017}%
  \BibitemOpen
  \bibfield  {author} {\bibinfo {author} {\bibfnamefont {J.~H.}\ \bibnamefont
  {Bauer}},\ }\bibfield  {title} {\bibinfo {title} {Search for the best
  analytical formula in the tunneling and the barrier-suppression ionization
  regimes},\ }\href {https://doi.org/10.1088/1361-6455/aa65ae} {\bibfield
  {journal} {\bibinfo  {journal} {J. Phys. B.}\ }\textbf {\bibinfo {volume}
  {50}},\ \bibinfo {pages} {085601} (\bibinfo {year} {2017})}\BibitemShut
  {NoStop}%
\bibitem [{\citenamefont {Remme}\ \emph {et~al.}(2025)\citenamefont {Remme},
  \citenamefont {Voitkiv}, \citenamefont {Pretzler},\ and\ \citenamefont
  {Müller}}]{Remme2025}%
  \BibitemOpen
  \bibfield  {author} {\bibinfo {author} {\bibfnamefont {S.}~\bibnamefont
  {Remme}}, \bibinfo {author} {\bibfnamefont {A.~B.}\ \bibnamefont {Voitkiv}},
  \bibinfo {author} {\bibfnamefont {G.}~\bibnamefont {Pretzler}},\ and\
  \bibinfo {author} {\bibfnamefont {C.}~\bibnamefont {Müller}},\ }\bibfield
  {title} {\bibinfo {title} {Phenomenological rate formulas for over-barrier
  ionization of hydrogen and helium atoms in strong constant electric fields},\
  }\href {https://arxiv.org/abs/2502.02697} {\bibfield  {journal} {\bibinfo
  {journal} {arXiv:2502.02697}\ } (\bibinfo {year} {2025})}\BibitemShut
  {NoStop}%
\bibitem [{\citenamefont {Yang}\ \emph {et~al.}(2012)\citenamefont {Yang},
  \citenamefont {Song},\ and\ \citenamefont {Chen}}]{Yang2012}%
  \BibitemOpen
  \bibfield  {author} {\bibinfo {author} {\bibfnamefont {W.}~\bibnamefont
  {Yang}}, \bibinfo {author} {\bibfnamefont {X.}~\bibnamefont {Song}},\ and\
  \bibinfo {author} {\bibfnamefont {Z.}~\bibnamefont {Chen}},\ }\bibfield
  {title} {\bibinfo {title} {Phase-dependent above-barrier ionization of
  excited-state electrons},\ }\href {https://doi.org/10.1364/OE.20.012067}
  {\bibfield  {journal} {\bibinfo  {journal} {Opt. Express}\ }\textbf {\bibinfo
  {volume} {20}},\ \bibinfo {pages} {12067} (\bibinfo {year}
  {2012})}\BibitemShut {NoStop}%
\bibitem [{\citenamefont {Bauer}\ \emph {et~al.}(2014)\citenamefont {Bauer},
  \citenamefont {Mota-Furtado}, \citenamefont {O'Mahony}, \citenamefont
  {Piraux},\ and\ \citenamefont {Warda}}]{Bauer2014}%
  \BibitemOpen
  \bibfield  {author} {\bibinfo {author} {\bibfnamefont {J.~H.}\ \bibnamefont
  {Bauer}}, \bibinfo {author} {\bibfnamefont {F.}~\bibnamefont {Mota-Furtado}},
  \bibinfo {author} {\bibfnamefont {P.~F.}\ \bibnamefont {O'Mahony}}, \bibinfo
  {author} {\bibfnamefont {B.}~\bibnamefont {Piraux}},\ and\ \bibinfo {author}
  {\bibfnamefont {K.}~\bibnamefont {Warda}},\ }\bibfield  {title} {\bibinfo
  {title} {Ionization and excitation of the excited hydrogen atom in strong
  circularly polarized laser fields},\ }\href
  {https://doi.org/10.1103/PhysRevA.90.063402} {\bibfield  {journal} {\bibinfo
  {journal} {Phys. Rev. A}\ }\textbf {\bibinfo {volume} {90}},\ \bibinfo
  {pages} {063402} (\bibinfo {year} {2014})}\BibitemShut {NoStop}%
\bibitem [{\citenamefont {Song}\ \emph {et~al.}(2014)\citenamefont {Song},
  \citenamefont {Geng}, \citenamefont {Jiang},\ and\ \citenamefont
  {Peng}}]{Song2014}%
  \BibitemOpen
  \bibfield  {author} {\bibinfo {author} {\bibfnamefont {S.-N.}\ \bibnamefont
  {Song}}, \bibinfo {author} {\bibfnamefont {J.-W.}\ \bibnamefont {Geng}},
  \bibinfo {author} {\bibfnamefont {H.-B.}\ \bibnamefont {Jiang}},\ and\
  \bibinfo {author} {\bibfnamefont {L.-Y.}\ \bibnamefont {Peng}},\ }\bibfield
  {title} {\bibinfo {title} {Comparative study of $\text{H}(2s)$ and
  $\text{H}(2{p}_{0})$ ionization dynamics in the over-barrier regime},\ }\href
  {https://doi.org/10.1103/PhysRevA.89.053411} {\bibfield  {journal} {\bibinfo
  {journal} {Phys. Rev. A}\ }\textbf {\bibinfo {volume} {89}},\ \bibinfo
  {pages} {053411} (\bibinfo {year} {2014})}\BibitemShut {NoStop}%
\bibitem [{\citenamefont {Jheng}\ \emph {et~al.}(2018)\citenamefont {Jheng},
  \citenamefont {Jiang}, \citenamefont {Chen},\ and\ \citenamefont
  {Liu}}]{Jheng2018}%
  \BibitemOpen
  \bibfield  {author} {\bibinfo {author} {\bibfnamefont {S.-D.}\ \bibnamefont
  {Jheng}}, \bibinfo {author} {\bibfnamefont {T.-F.}\ \bibnamefont {Jiang}},
  \bibinfo {author} {\bibfnamefont {J.-H.}\ \bibnamefont {Chen}},\ and\
  \bibinfo {author} {\bibfnamefont {J.-L.}\ \bibnamefont {Liu}},\ }\bibfield
  {title} {\bibinfo {title} {Magnetic quantum number dependence of hydrogen
  photoelectron spectra under circularly polarized pulse in barrier suppression
  ionization regime},\ }\href {https://doi.org/10.1088/1402-4896/aaca08}
  {\bibfield  {journal} {\bibinfo  {journal} {Phys. Scr.}\ }\textbf {\bibinfo
  {volume} {93}},\ \bibinfo {pages} {085401} (\bibinfo {year}
  {2018})}\BibitemShut {NoStop}%
\bibitem [{\citenamefont {Belsa}\ \emph {et~al.}(2022)\citenamefont {Belsa},
  \citenamefont {Ziems}, \citenamefont {Sanchez}, \citenamefont {Chirvi},
  \citenamefont {Liu}, \citenamefont {Gr\"afe},\ and\ \citenamefont
  {Biegert}}]{Belsa2022}%
  \BibitemOpen
  \bibfield  {author} {\bibinfo {author} {\bibfnamefont {B.}~\bibnamefont
  {Belsa}}, \bibinfo {author} {\bibfnamefont {K.~M.}\ \bibnamefont {Ziems}},
  \bibinfo {author} {\bibfnamefont {A.}~\bibnamefont {Sanchez}}, \bibinfo
  {author} {\bibfnamefont {K.}~\bibnamefont {Chirvi}}, \bibinfo {author}
  {\bibfnamefont {X.}~\bibnamefont {Liu}}, \bibinfo {author} {\bibfnamefont
  {S.}~\bibnamefont {Gr\"afe}},\ and\ \bibinfo {author} {\bibfnamefont
  {J.}~\bibnamefont {Biegert}},\ }\bibfield  {title} {\bibinfo {title}
  {Laser-induced electron diffraction in the over-the-barrier-ionization
  regime},\ }\href {https://doi.org/10.1103/PhysRevA.106.043105} {\bibfield
  {journal} {\bibinfo  {journal} {Phys. Rev. A}\ }\textbf {\bibinfo {volume}
  {106}},\ \bibinfo {pages} {043105} (\bibinfo {year} {2022})}\BibitemShut
  {NoStop}%
\bibitem [{\citenamefont {Ma}\ \emph {et~al.}(2023)\citenamefont {Ma},
  \citenamefont {Wang}, \citenamefont {Zhang}, \citenamefont {Zou},
  \citenamefont {Yuan}, \citenamefont {Ma}, \citenamefont {Lv}, \citenamefont
  {Shen}, \citenamefont {Yan}, \citenamefont {Weidem\"uller}, \citenamefont
  {Ye},\ and\ \citenamefont {Jiang}}]{Ma2023}%
  \BibitemOpen
  \bibfield  {author} {\bibinfo {author} {\bibfnamefont {H.}~\bibnamefont
  {Ma}}, \bibinfo {author} {\bibfnamefont {X.}~\bibnamefont {Wang}}, \bibinfo
  {author} {\bibfnamefont {L.}~\bibnamefont {Zhang}}, \bibinfo {author}
  {\bibfnamefont {Z.}~\bibnamefont {Zou}}, \bibinfo {author} {\bibfnamefont
  {J.}~\bibnamefont {Yuan}}, \bibinfo {author} {\bibfnamefont {Y.}~\bibnamefont
  {Ma}}, \bibinfo {author} {\bibfnamefont {R.}~\bibnamefont {Lv}}, \bibinfo
  {author} {\bibfnamefont {Z.}~\bibnamefont {Shen}}, \bibinfo {author}
  {\bibfnamefont {T.}~\bibnamefont {Yan}}, \bibinfo {author} {\bibfnamefont
  {M.}~\bibnamefont {Weidem\"uller}}, \bibinfo {author} {\bibfnamefont
  {D.}~\bibnamefont {Ye}},\ and\ \bibinfo {author} {\bibfnamefont
  {Y.}~\bibnamefont {Jiang}},\ }\bibfield  {title} {\bibinfo {title}
  {Few-photon single ionization of cold rubidium in the over-the-barrier
  regime},\ }\href {https://doi.org/10.1103/PhysRevA.107.033114} {\bibfield
  {journal} {\bibinfo  {journal} {Phys. Rev. A}\ }\textbf {\bibinfo {volume}
  {107}},\ \bibinfo {pages} {033114} (\bibinfo {year} {2023})}\BibitemShut
  {NoStop}%
\bibitem [{\citenamefont {Ni}\ \emph {et~al.}(2016)\citenamefont {Ni},
  \citenamefont {Saalmann},\ and\ \citenamefont {Rost}}]{Ni2016}%
  \BibitemOpen
  \bibfield  {author} {\bibinfo {author} {\bibfnamefont {H.}~\bibnamefont
  {Ni}}, \bibinfo {author} {\bibfnamefont {U.}~\bibnamefont {Saalmann}},\ and\
  \bibinfo {author} {\bibfnamefont {J.-M.}\ \bibnamefont {Rost}},\ }\bibfield
  {title} {\bibinfo {title} {Tunneling ionization time resolved by
  backpropagation},\ }\href {https://doi.org/10.1103/PhysRevLett.117.023002}
  {\bibfield  {journal} {\bibinfo  {journal} {Phys. Rev. Lett.}\ }\textbf
  {\bibinfo {volume} {117}},\ \bibinfo {pages} {023002} (\bibinfo {year}
  {2016})}\BibitemShut {NoStop}%
\bibitem [{\citenamefont {Ni}\ \emph {et~al.}(2018{\natexlab{a}})\citenamefont
  {Ni}, \citenamefont {Saalmann},\ and\ \citenamefont {Rost}}]{Ni2018}%
  \BibitemOpen
  \bibfield  {author} {\bibinfo {author} {\bibfnamefont {H.}~\bibnamefont
  {Ni}}, \bibinfo {author} {\bibfnamefont {U.}~\bibnamefont {Saalmann}},\ and\
  \bibinfo {author} {\bibfnamefont {J.-M.}\ \bibnamefont {Rost}},\ }\bibfield
  {title} {\bibinfo {title} {Tunneling exit characteristics from classical
  backpropagation of an ionized electron wave packet},\ }\href
  {https://doi.org/10.1103/PhysRevA.97.013426} {\bibfield  {journal} {\bibinfo
  {journal} {Phys. Rev. A}\ }\textbf {\bibinfo {volume} {97}},\ \bibinfo
  {pages} {013426} (\bibinfo {year} {2018}{\natexlab{a}})}\BibitemShut
  {NoStop}%
\bibitem [{\citenamefont {Ni}\ \emph {et~al.}(2018{\natexlab{b}})\citenamefont
  {Ni}, \citenamefont {Eicke}, \citenamefont {Ruiz}, \citenamefont {Cai},
  \citenamefont {Oppermann}, \citenamefont {Shvetsov-Shilovski},\ and\
  \citenamefont {Pi}}]{Ni2018a}%
  \BibitemOpen
  \bibfield  {author} {\bibinfo {author} {\bibfnamefont {H.}~\bibnamefont
  {Ni}}, \bibinfo {author} {\bibfnamefont {N.}~\bibnamefont {Eicke}}, \bibinfo
  {author} {\bibfnamefont {C.}~\bibnamefont {Ruiz}}, \bibinfo {author}
  {\bibfnamefont {J.}~\bibnamefont {Cai}}, \bibinfo {author} {\bibfnamefont
  {F.}~\bibnamefont {Oppermann}}, \bibinfo {author} {\bibfnamefont {N.~I.}\
  \bibnamefont {Shvetsov-Shilovski}},\ and\ \bibinfo {author} {\bibfnamefont
  {L.-W.}\ \bibnamefont {Pi}},\ }\bibfield  {title} {\bibinfo {title}
  {Tunneling criteria and a nonadiabatic term for strong-field ionization},\
  }\href {https://doi.org/10.1103/PhysRevA.98.013411} {\bibfield  {journal}
  {\bibinfo  {journal} {Phys. Rev. A}\ }\textbf {\bibinfo {volume} {98}},\
  \bibinfo {pages} {013411} (\bibinfo {year} {2018}{\natexlab{b}})}\BibitemShut
  {NoStop}%
\bibitem [{\citenamefont {Hofmann}\ \emph {et~al.}(2021)\citenamefont
  {Hofmann}, \citenamefont {Bray}, \citenamefont {Koch}, \citenamefont {Ni},\
  and\ \citenamefont {Shvetsov-Shilovski}}]{Hofmann2021}%
  \BibitemOpen
  \bibfield  {author} {\bibinfo {author} {\bibfnamefont {C.}~\bibnamefont
  {Hofmann}}, \bibinfo {author} {\bibfnamefont {A.}~\bibnamefont {Bray}},
  \bibinfo {author} {\bibfnamefont {W.}~\bibnamefont {Koch}}, \bibinfo {author}
  {\bibfnamefont {H.}~\bibnamefont {Ni}},\ and\ \bibinfo {author}
  {\bibfnamefont {N.~I.}\ \bibnamefont {Shvetsov-Shilovski}},\ }\bibfield
  {title} {\bibinfo {title} {{Quantum battles in attoscience: tunnelling}},\
  }\href {https://doi.org/10.1140/epjd/s10053-021-00224-2} {\bibfield
  {journal} {\bibinfo  {journal} {Eur. Phys. J. D}\ }\textbf {\bibinfo {volume}
  {75}},\ \bibinfo {pages} {208} (\bibinfo {year} {2021})}\BibitemShut
  {NoStop}%
\bibitem [{\citenamefont {Ni}\ \emph {et~al.}(2020)\citenamefont {Ni},
  \citenamefont {Brennecke}, \citenamefont {Gao}, \citenamefont {He},
  \citenamefont {Donsa}, \citenamefont {B{\v{r}}ezinov{\'a}}, \citenamefont
  {He}, \citenamefont {Wu}, \citenamefont {Lein}, \citenamefont {Tong},\ and\
  \citenamefont {Burgd{\"o}rfer}}]{Ni2020}%
  \BibitemOpen
  \bibfield  {author} {\bibinfo {author} {\bibfnamefont {H.}~\bibnamefont
  {Ni}}, \bibinfo {author} {\bibfnamefont {S.}~\bibnamefont {Brennecke}},
  \bibinfo {author} {\bibfnamefont {X.}~\bibnamefont {Gao}}, \bibinfo {author}
  {\bibfnamefont {P.-L.}\ \bibnamefont {He}}, \bibinfo {author} {\bibfnamefont
  {S.}~\bibnamefont {Donsa}}, \bibinfo {author} {\bibfnamefont
  {I.}~\bibnamefont {B{\v{r}}ezinov{\'a}}}, \bibinfo {author} {\bibfnamefont
  {F.}~\bibnamefont {He}}, \bibinfo {author} {\bibfnamefont {J.}~\bibnamefont
  {Wu}}, \bibinfo {author} {\bibfnamefont {M.}~\bibnamefont {Lein}}, \bibinfo
  {author} {\bibfnamefont {X.-M.}\ \bibnamefont {Tong}},\ and\ \bibinfo
  {author} {\bibfnamefont {J.}~\bibnamefont {Burgd{\"o}rfer}},\ }\bibfield
  {title} {\bibinfo {title} {Theory of subcycle linear momentum transfer in
  strong-field tunneling ionization},\ }\href
  {https://doi.org/10.1103/PhysRevLett.125.073202} {\bibfield  {journal}
  {\bibinfo  {journal} {Phys. Rev. Lett.}\ }\textbf {\bibinfo {volume} {125}},\
  \bibinfo {pages} {073202} (\bibinfo {year} {2020})}\BibitemShut {NoStop}%
\bibitem [{\citenamefont {Kim}\ \emph {et~al.}(2021)\citenamefont {Kim},
  \citenamefont {Ivanov},\ and\ \citenamefont {Kim}}]{Kim2021}%
  \BibitemOpen
  \bibfield  {author} {\bibinfo {author} {\bibfnamefont {Y.~H.}\ \bibnamefont
  {Kim}}, \bibinfo {author} {\bibfnamefont {I.~A.}\ \bibnamefont {Ivanov}},\
  and\ \bibinfo {author} {\bibfnamefont {K.~T.}\ \bibnamefont {Kim}},\
  }\bibfield  {title} {\bibinfo {title} {{Classical backpropagation for probing
  the backward rescattering time of a tunnel-ionized electron in an intense
  laser field}},\ }\href {https://doi.org/10.1103/PhysRevA.104.013116}
  {\bibfield  {journal} {\bibinfo  {journal} {Phys. Rev. A}\ }\textbf {\bibinfo
  {volume} {104}},\ \bibinfo {pages} {013116} (\bibinfo {year}
  {2021})}\BibitemShut {NoStop}%
\bibitem [{\citenamefont {Ma}\ \emph {et~al.}(2024{\natexlab{b}})\citenamefont
  {Ma}, \citenamefont {Ni}, \citenamefont {Li}, \citenamefont {He},\ and\
  \citenamefont {Wu}}]{Ma2024US}%
  \BibitemOpen
  \bibfield  {author} {\bibinfo {author} {\bibfnamefont {Y.}~\bibnamefont
  {Ma}}, \bibinfo {author} {\bibfnamefont {H.}~\bibnamefont {Ni}}, \bibinfo
  {author} {\bibfnamefont {Y.}~\bibnamefont {Li}}, \bibinfo {author}
  {\bibfnamefont {F.}~\bibnamefont {He}},\ and\ \bibinfo {author}
  {\bibfnamefont {J.}~\bibnamefont {Wu}},\ }\bibfield  {title} {\bibinfo
  {title} {{Subcycle Conservation Law in Strong-Field Ionization}},\ }\href
  {https://doi.org/10.34133/ultrafastscience.0071} {\bibfield  {journal}
  {\bibinfo  {journal} {Ultrafast Sci.}\ }\textbf {\bibinfo {volume} {4}},\
  \bibinfo {pages} {0071} (\bibinfo {year} {2024}{\natexlab{b}})}\BibitemShut
  {NoStop}%
\bibitem [{\citenamefont {Klaiber}\ \emph {et~al.}(2013)\citenamefont
  {Klaiber}, \citenamefont {Yakaboylu}, \citenamefont {Bauke}, \citenamefont
  {Hatsagortsyan},\ and\ \citenamefont {Keitel}}]{Klaiber2013}%
  \BibitemOpen
  \bibfield  {author} {\bibinfo {author} {\bibfnamefont {M.}~\bibnamefont
  {Klaiber}}, \bibinfo {author} {\bibfnamefont {E.}~\bibnamefont {Yakaboylu}},
  \bibinfo {author} {\bibfnamefont {H.}~\bibnamefont {Bauke}}, \bibinfo
  {author} {\bibfnamefont {K.~Z.}\ \bibnamefont {Hatsagortsyan}},\ and\
  \bibinfo {author} {\bibfnamefont {C.~H.}\ \bibnamefont {Keitel}},\ }\bibfield
   {title} {\bibinfo {title} {Under-the-barrier dynamics in laser-induced
  relativistic tunneling},\ }\href
  {https://doi.org/10.1103/PhysRevLett.110.153004} {\bibfield  {journal}
  {\bibinfo  {journal} {Phys. Rev. Lett.}\ }\textbf {\bibinfo {volume} {110}},\
  \bibinfo {pages} {153004} (\bibinfo {year} {2013})}\BibitemShut {NoStop}%
\bibitem [{\citenamefont {Li}\ \emph {et~al.}(2016)\citenamefont {Li},
  \citenamefont {Geng}, \citenamefont {Han}, \citenamefont {Liu}, \citenamefont
  {Peng}, \citenamefont {Gong},\ and\ \citenamefont {Liu}}]{Li2016}%
  \BibitemOpen
  \bibfield  {author} {\bibinfo {author} {\bibfnamefont {M.}~\bibnamefont
  {Li}}, \bibinfo {author} {\bibfnamefont {J.-W.}\ \bibnamefont {Geng}},
  \bibinfo {author} {\bibfnamefont {M.}~\bibnamefont {Han}}, \bibinfo {author}
  {\bibfnamefont {M.-M.}\ \bibnamefont {Liu}}, \bibinfo {author} {\bibfnamefont
  {L.-Y.}\ \bibnamefont {Peng}}, \bibinfo {author} {\bibfnamefont
  {Q.}~\bibnamefont {Gong}},\ and\ \bibinfo {author} {\bibfnamefont
  {Y.}~\bibnamefont {Liu}},\ }\bibfield  {title} {\bibinfo {title} {Subcycle
  nonadiabatic strong-field tunneling ionization},\ }\href
  {https://doi.org/10.1103/PhysRevA.93.013402} {\bibfield  {journal} {\bibinfo
  {journal} {Phys. Rev. A}\ }\textbf {\bibinfo {volume} {93}},\ \bibinfo
  {pages} {013402} (\bibinfo {year} {2016})}\BibitemShut {NoStop}%
\bibitem [{\citenamefont {Eckart}\ \emph {et~al.}(2018)\citenamefont {Eckart},
  \citenamefont {Fehre}, \citenamefont {Eicke}, \citenamefont {Hartung},
  \citenamefont {Rist}, \citenamefont {Trabert}, \citenamefont {Strenger},
  \citenamefont {Pier}, \citenamefont {Schmidt}, \citenamefont {Jahnke},
  \citenamefont {Sch\"offler}, \citenamefont {Lein}, \citenamefont {Kunitski},\
  and\ \citenamefont {D\"orner}}]{Eckart2018}%
  \BibitemOpen
  \bibfield  {author} {\bibinfo {author} {\bibfnamefont {S.}~\bibnamefont
  {Eckart}}, \bibinfo {author} {\bibfnamefont {K.}~\bibnamefont {Fehre}},
  \bibinfo {author} {\bibfnamefont {N.}~\bibnamefont {Eicke}}, \bibinfo
  {author} {\bibfnamefont {A.}~\bibnamefont {Hartung}}, \bibinfo {author}
  {\bibfnamefont {J.}~\bibnamefont {Rist}}, \bibinfo {author} {\bibfnamefont
  {D.}~\bibnamefont {Trabert}}, \bibinfo {author} {\bibfnamefont
  {N.}~\bibnamefont {Strenger}}, \bibinfo {author} {\bibfnamefont
  {A.}~\bibnamefont {Pier}}, \bibinfo {author} {\bibfnamefont {L.~P.~H.}\
  \bibnamefont {Schmidt}}, \bibinfo {author} {\bibfnamefont {T.}~\bibnamefont
  {Jahnke}}, \bibinfo {author} {\bibfnamefont {M.~S.}\ \bibnamefont
  {Sch\"offler}}, \bibinfo {author} {\bibfnamefont {M.}~\bibnamefont {Lein}},
  \bibinfo {author} {\bibfnamefont {M.}~\bibnamefont {Kunitski}},\ and\
  \bibinfo {author} {\bibfnamefont {R.}~\bibnamefont {D\"orner}},\ }\bibfield
  {title} {\bibinfo {title} {Direct experimental access to the nonadiabatic
  initial momentum offset upon tunnel ionization},\ }\href
  {https://doi.org/10.1103/PhysRevLett.121.163202} {\bibfield  {journal}
  {\bibinfo  {journal} {Phys. Rev. Lett.}\ }\textbf {\bibinfo {volume} {121}},\
  \bibinfo {pages} {163202} (\bibinfo {year} {2018})}\BibitemShut {NoStop}%
\bibitem [{\citenamefont {Liu}\ \emph {et~al.}(2019)\citenamefont {Liu},
  \citenamefont {Luo}, \citenamefont {Li}, \citenamefont {Li}, \citenamefont
  {Feng}, \citenamefont {Du}, \citenamefont {Zhou}, \citenamefont {Lu},\ and\
  \citenamefont {Barth}}]{Liu2019}%
  \BibitemOpen
  \bibfield  {author} {\bibinfo {author} {\bibfnamefont {K.}~\bibnamefont
  {Liu}}, \bibinfo {author} {\bibfnamefont {S.}~\bibnamefont {Luo}}, \bibinfo
  {author} {\bibfnamefont {M.}~\bibnamefont {Li}}, \bibinfo {author}
  {\bibfnamefont {Y.}~\bibnamefont {Li}}, \bibinfo {author} {\bibfnamefont
  {Y.}~\bibnamefont {Feng}}, \bibinfo {author} {\bibfnamefont {B.}~\bibnamefont
  {Du}}, \bibinfo {author} {\bibfnamefont {Y.}~\bibnamefont {Zhou}}, \bibinfo
  {author} {\bibfnamefont {P.}~\bibnamefont {Lu}},\ and\ \bibinfo {author}
  {\bibfnamefont {I.}~\bibnamefont {Barth}},\ }\bibfield  {title} {\bibinfo
  {title} {Detecting and characterizing the nonadiabaticity of laser-induced
  quantum tunneling},\ }\href {https://doi.org/10.1103/PhysRevLett.122.053202}
  {\bibfield  {journal} {\bibinfo  {journal} {Phys. Rev. Lett.}\ }\textbf
  {\bibinfo {volume} {122}},\ \bibinfo {pages} {053202} (\bibinfo {year}
  {2019})}\BibitemShut {NoStop}%
\bibitem [{\citenamefont {Klaiber}\ \emph {et~al.}(2022)\citenamefont
  {Klaiber}, \citenamefont {Lv}, \citenamefont {Sukiasyan}, \citenamefont
  {Bakucz~Can\'ario}, \citenamefont {Hatsagortsyan},\ and\ \citenamefont
  {Keitel}}]{Klaiber2022}%
  \BibitemOpen
  \bibfield  {author} {\bibinfo {author} {\bibfnamefont {M.}~\bibnamefont
  {Klaiber}}, \bibinfo {author} {\bibfnamefont {Q.~Z.}\ \bibnamefont {Lv}},
  \bibinfo {author} {\bibfnamefont {S.}~\bibnamefont {Sukiasyan}}, \bibinfo
  {author} {\bibfnamefont {D.}~\bibnamefont {Bakucz~Can\'ario}}, \bibinfo
  {author} {\bibfnamefont {K.~Z.}\ \bibnamefont {Hatsagortsyan}},\ and\
  \bibinfo {author} {\bibfnamefont {C.~H.}\ \bibnamefont {Keitel}},\ }\bibfield
   {title} {\bibinfo {title} {Reconciling conflicting approaches for the
  tunneling time delay in strong field ionization},\ }\href
  {https://doi.org/10.1103/PhysRevLett.129.203201} {\bibfield  {journal}
  {\bibinfo  {journal} {Phys. Rev. Lett.}\ }\textbf {\bibinfo {volume} {129}},\
  \bibinfo {pages} {203201} (\bibinfo {year} {2022})}\BibitemShut {NoStop}%
\bibitem [{\citenamefont {van~de Sand}\ and\ \citenamefont
  {Rost}(1999)}]{vandeSand1999}%
  \BibitemOpen
  \bibfield  {author} {\bibinfo {author} {\bibfnamefont {G.}~\bibnamefont
  {van~de Sand}}\ and\ \bibinfo {author} {\bibfnamefont {J.~M.}\ \bibnamefont
  {Rost}},\ }\bibfield  {title} {\bibinfo {title} {Irregular orbits generate
  higher harmonics},\ }\href {https://doi.org/10.1103/PhysRevLett.83.524}
  {\bibfield  {journal} {\bibinfo  {journal} {Phys. Rev. Lett.}\ }\textbf
  {\bibinfo {volume} {83}},\ \bibinfo {pages} {524} (\bibinfo {year}
  {1999})}\BibitemShut {NoStop}%
\bibitem [{\citenamefont {Rodr\'{\i}guez-Hern\'andez}\ \emph
  {et~al.}(2022)\citenamefont {Rodr\'{\i}guez-Hern\'andez}, \citenamefont
  {Grossmann},\ and\ \citenamefont {Rost}}]{Rodriguez2022}%
  \BibitemOpen
  \bibfield  {author} {\bibinfo {author} {\bibfnamefont {F.}~\bibnamefont
  {Rodr\'{\i}guez-Hern\'andez}}, \bibinfo {author} {\bibfnamefont
  {F.}~\bibnamefont {Grossmann}},\ and\ \bibinfo {author} {\bibfnamefont
  {J.~M.}\ \bibnamefont {Rost}},\ }\bibfield  {title} {\bibinfo {title} {Cutoff
  for continuum-continuum high-order harmonic generation},\ }\href
  {https://doi.org/10.1103/PhysRevA.105.L051102} {\bibfield  {journal}
  {\bibinfo  {journal} {Phys. Rev. A}\ }\textbf {\bibinfo {volume} {105}},\
  \bibinfo {pages} {L051102} (\bibinfo {year} {2022})}\BibitemShut {NoStop}%
\end{thebibliography}%

\end{document}